\documentclass[twocolumn,amsmath,amssymb,prapplied,longbibliography]{revtex4-2}

\usepackage{graphicx}
\usepackage{dcolumn}
\usepackage{bm}
\usepackage{color}
\usepackage{notes2bib}

\begin{document}



\title{Engineering chlorine-based emitters in silicon carbide for telecom-band quantum technologies}

\author{A.~N.~Anisimov$^{1,3}$}
\email[E-mail:~]{a.anisimov@hzdr.de}
\author{A.~V.~Mathews$^{1,2,3}$}
\author{K.~Mavridou$^{1}$}
\author{U.~Kentsch$^{1}$}
\author{M.~Helm$^{1,2}$}
\author{G.~V.~Astakhov$^{1}$}
\email[E-mail:~]{g.astakhov@hzdr.de}

\affiliation{$^1$Helmholtz-Zentrum Dresden-Rossendorf, Institute of Ion Beam Physics and Materials Research, 01328 Dresden, Germany  \\
$^2$Technische Universit\"{a}t Dresden, 01062 Dresden, Germany \\
$^3$These authors contributed equally to this work
 }

\begin{abstract}
We report the experimental realization and optical characterization of chlorine-vacancy (ClV) color centers in 4H-SiC emitting in the fiber-optic telecom bands. These defects are created via chlorine ion implantation followed by high-temperature annealing. Photoluminescence spectroscopy reveals four distinct ClV configurations with zero-phonon lines located in the O-band  ($1260 - 1360 \, \mathrm{nm}$), S-band  ($1460 - 1530 \, \mathrm{nm}$) and C-band  ($1530 - 1565 \, \mathrm{nm}$). Controlled implantation and annealing experiments confirm that the ClV centers originate specifically from chlorine incorporation into SiC and are not intrinsic to this material. We optimize the creation conditions for ClV ensembles and demonstrate negligible reduction of the ZPL intensity up to a temperature of $30 \, \mathrm{K}$. These results establish ClV defects as a new class of telecom-band color centers in a CMOS-compatible platform, offering strong potential for scalable quantum networks.
\end{abstract}

\date{\today}

\maketitle

\section{Introduction}

Quantum networks, in which quantum registers are interconnected through optical fibers, provide a scalable architecture for the distribution and processing of quantum information \cite{10.1038/nature07127}. They enable essential functionalities of quantum technologies, including secure communication \cite{10.1126/science.aam9288}, distributed quantum computation \cite{10.1103/physreva.89.022317} and remote quantum sensing \cite{10.1002/lpor.201900097}. Optically active atom-like defects, or color centers, in solid-state materials offer a promising platform for the experimental realization of such networks. A prominent example is the nitrogen-vacancy (NV) center in diamond \cite{10.1063/5.0056534} with well-established spin-photon interface \cite{10.1038/nature15759} and outstanding coherence properties  \cite{10.1038/nmat2420}. Using the NV or certain group-IV defects in diamond, such as the silicon-vacancy (SiV), multi-qubit quantum network nodes have already been experimentally demonstrated \cite{10.1126/science.abg1919, 10.1126/science.add9771}. However, these color centers emit in the visible spectral range, which hampers direct integration with low-loss fiber networks and necessitates quantum frequency conversion into the telecom band  \cite{10.1103/PhysRevLett.123.063601}.  This challenge motivates the search of telecom-wavelength quantum emitters in scalable material platforms such as silicon and silicon carbide (SiC).

Indeed, telecom single-photon emitters (SPEs) based on the G-center have been isolated in silicon \cite{10.1364/oe.397377, 10.1038/s41928-020-00499-0}. While the G-center is among the brightest SPEs in the telecom O-band, its non- zero spin resides only in a metastable state  \cite{10.1103/physrevlett.127.196402}, which precludes its use for robust spin-photon entanglement and, therefore, may limit its suitability for quantum networking applications. The only known SPEs in silicon with emission in the telecom band and a demonstrated spin-photon interface are the T-centers \cite{10.1038/s41586-022-04821-y} and erbium (Er) dopants \cite{10.1038/s41467-024-55552-9}. However, both suffer from low brightness due to inherently long radiative lifetimes.

Silicon carbide (SiC) is another attractive material platform for scalable quantum networks due to its wide bandgap, mature wafer-scale processing and compatibility with CMOS infrastructure  \cite{10.1063/5.0004454, 10.1063/5.0262377}.  The most extensively studied defects in SiC for spin-photonic quantum applications are silicon vacancies ($\mathrm{V_{Si}}$) \cite{10.1103/physrevlett.109.226402} and divacancies \cite{10.1038/nature10562}, which feature spin-selective optical transitions \cite{10.1038/nphys2826, 10.1126/sciadv.1501015}, long spin coherence times  \cite{10.1103/physrevb.95.161201} as well as the ability to be individually addressed and coherently controlled  \cite{10.1038/nmat4145, 10.1038/nmat4144, 10.1038/ncomms8578, 10.1038/s41467-019-09873-9}. In addition, the NV centers in SiC have attracted significant interest due to their strong analogies to the NV center in diamond, but within a more technologically mature and scalable host \cite{10.1103/physrevb.94.060102, 10.1103/physrevlett.124.223601}. All these color centers can also be  integrated into nanophotonic structures, enabling Purcell enhancement and efficient photon extraction  \cite{10.1021/acs.nanolett.6b05102, 10.1038/s41563-021-01148-3, 10.1021/acs.nanolett.0c00339, 10.1021/acsphotonics.5c00096}, which are key requirements for scalable quantum photonic networks.

While $\mathrm{V_{Si}}$, divacancies and NV in SiC emit in the near-infrared spectral range ($800 - 1200 \, \mathrm{nm}$), recent efforts have focused on searching and engineering color centers in SiC that operate directly in the telecom bands (O, E, S, C, L, U) \cite{10.1002/qute.202100076}.  Particularly, Er dopants can also be implanted into SiC, exhibiting emission around $1550 \, \mathrm{nm}$ (C-band) \cite{10.1063/5.0055100}. However, their experimentally demonstrated optical properties remain relatively limited. Currently, the most promising color center in SiC for telecom applications is vanadium (V)  \cite{10.1088/2058-9565/ad48b1}, which exhibits emission in the O-band, specifically with the zero-phonon lines (ZPLs) around $1278 \, \mathrm{nm}$ and $1334 \,\mathrm{nm}$ in 4H-SiC for the $\mathrm{V (\alpha)}$ and $\mathrm{V (\beta)}$ configurations, respectively  \cite{10.1103/physrevapplied.12.014015}. Coherent control of single vanadium centers has been demonstrated \cite{10.1126/sciadv.aaz1192}, together with extended spin relaxation times \cite{10.1103/physrevapplied.22.044078} and ultra-narrow inhomogeneous spectral distributions in isotopically purified SiC \cite{10.1038/s41467-023-43923-7}.  In comparison with $\mathrm{V_{Si}}$ and divacancies under similar conditions, the V color centers exhibit significantly lower optical transition rates and their coherent control schemes are more complex, requiring sub-Kelvin temperatures \cite{10.1103/physrevapplied.22.044078}. These limitations motivate the theoretical search for alternative, i.e.,  NV-like color centers in SiC but with emission in the telecom range.

Very recently, high-accuracy density functional theory (DFT) calculations have been used to investigate a variety of new point defects in SiC  \cite{10.1103/physrevb.108.224106}. Among them, the chlorine-vacancy (ClV) centers, where a Cl atom substitutes a C atom and coupled to a neighbour Si vacancy, have been predicted to exhibit optical and spin properties similar to the NV center in diamond, but with four ZPLs emitting in different telecom bands, ranging from $1330 \, \mathrm{nm}$ to $1590 \, \mathrm{nm}$ \cite{10.1103/physrevb.108.224106}. The formation of Cl-related complexes in $p$-type SiC has been inferred from deep-level transient spectroscopy measurements~\cite{10.4028/www.scientific.net/msf.717-720.229}. However, no optical emission associated with these complexes has been observed to date.

Here, we report the first optical observation of ClV centers in various SiC wafers, created by Cl implantation. Their optical emission consisting of ZPLs and phonon side bands (PSBs) spans the entire fiber-optic telecommunication window, confirmed by photoluminescence (PL) spectroscopy. Notably, the ClV ZPLs are located in  the O-, S- and C-bands, spectral regions of paramount importance for long-distance quantum communication. These properties establish ClV centers as a new class of SiC defects that are inherently compatible with existing telecom infrastructure and hold strong potential for scalable quantum technologies.

\section{Experiment}

\subsection{Fabrication protocol of ClV defects}

\begin{figure}[t]
\centering\includegraphics[width=.49\textwidth]{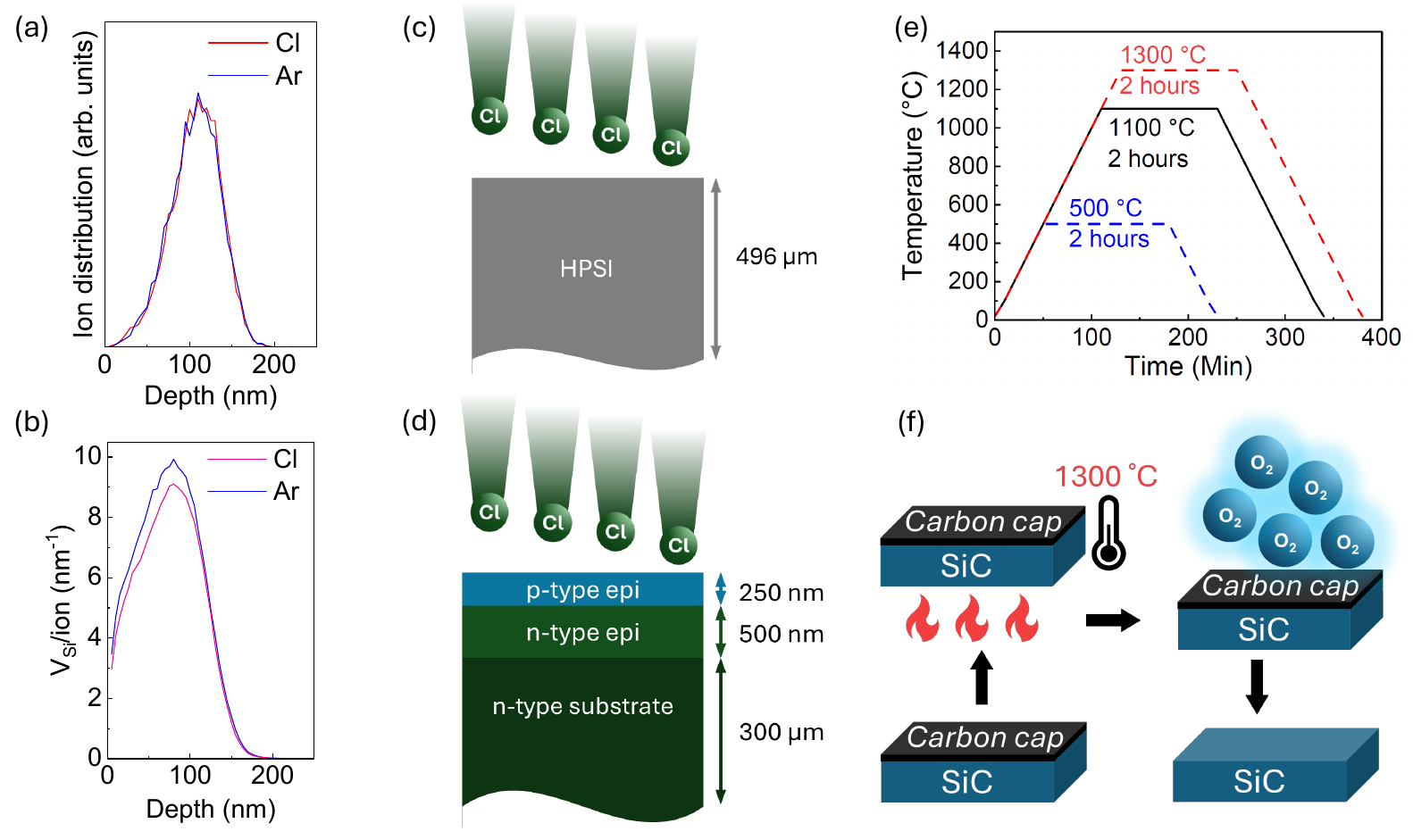}
\caption{Engineering of ClV defects in 4H-SiC. (a) In-depth distribution profile of Cl and Ar ions calculated for the implantation energy  $E = 150 \, \mathrm{keV}$. (b) SRIM-simulated probability of creating $\mathrm{V_{Si}}$ per single $\mathrm{Cl^{+}}$ and $\mathrm{Ar^{+}}$ ions. (c) Schematic representation of the $\mathrm{Cl}$ implantation into a HPSI 4H-SiC sample \#1 with a thickness of  $496 \, \mathrm{\mu m}$. (d)  Schematic representation of the $\mathrm{Cl}$ implantation into the 4H-SiC sample \#2 with epilayers. (e)  Temperature profile of the annealing process. The diagram illustrates the time-dependent temperature evolution, including heating and cooling phases, with annealing conducted at $500 ^{\circ} \mathrm{C}$ (dashed blue), $1100 ^{\circ} \mathrm{C}$  (black) and $1300 ^{\circ} \mathrm{C}$ (dashed red) over 2 hours. (f) Schematic illustration of 4H-SiC coating with a protective carbon cap layer. Arrows indicate the sequence of four key steps used to prevent surface degradation during vacuum annealing: carbon cap deposition, annealing at $1300 ^{\circ} \mathrm{C}$, cap layer removal in oxygen plasma and final surface cleaning.}  
\label{fig2}
\end{figure}

To create ClV defects in SiC, we implant singly-charged chlorine ions $\mathrm{Cl^{+}}$ (atomic number 17)  at energies  $E = 40 \, \mathrm{keV}$ and $E = 150 \, \mathrm{keV}$
with fluences ranging from $\Phi = 1 \times 10^{11}\,\mathrm{cm^{-2}}$ to $\Phi = 5 \times 10^{14}\,\mathrm{cm^{-2}}$. For comparison, we also perform implantation with  singly-charged argon ions $\mathrm{Ar^{+}}$ (atomic number 18). The Stopping and Range of Ions in Matter (SRIM) simulation \cite{10.1016/j.nimb.2010.02.091} of the Cl and Ar distribution for $E = 150 \, \mathrm{keV}$ is presented in Fig.~\ref{fig2}a, which are nearly identical with the maximum at around $110 \, \mathrm{nm}$ below the surface. The SRIM-simulated probability of creating $\mathrm{V_{Si}}$ per single $\mathrm{Cl^{+}}$ and $\mathrm{Ar^{+}}$ ion is presented in Fig.~\ref{fig2}b. They have very similar in-depth distribution with only 5\% higher probability for the maximum  at around $90 \, \mathrm{nm}$ for $\mathrm{Ar^{+}}$ compared to $\mathrm{Cl^{+}}$. 

We investigate three types of SiC samples, commercially sourced from Norstel AB (now part of STMicroelectronics). Sample \#1 is a high-purity semi-insulating (HPSI) 4H-SiC wafer with a thickness of $496 \, \mathrm{\mu m}$ (Fig.~\ref{fig2}c). The sample is nominally undoped. The concentration of intrinsic impurities is not specified and it has resistivity $\geq 1 \times 10^{7}\,\mathrm{\Omega \cdot cm}$ at room temperature. Sample \#2 consists of 4H-SiC layers epitaxially grown on a n-type substrate (Fig.~\ref{fig2}d). The top-layer is p-type doped with aluminium (Al) to a level of $1 \times 10^{17} \, \mathrm{cm ^{-3}}$ and has a thickness of $250 \, \mathrm{nm}$. The buffer layer is n-type doped with nitrogen (N) to a level of $1 \times 10^{18} \, \mathrm{cm ^{-3}}$ and has a thickness of $500 \, \mathrm{nm}$. Sample \#3 also consists of 4H-SiC layers epitaxially grown on a n-type substrate. But the top-layer is n-type doped with nitrogen (N) to a level of $1 \times 10^{15} \, \mathrm{cm ^{-3}}$ and has a thickness of $20 \, \mathrm{\mu m}$. The buffer layer is n-type doped with nitrogen (N) to a level of $1 \times 10^{18} \, \mathrm{cm ^{-3}}$ and has a thickness of $500 \, \mathrm{nm}$.  

To activate ClV centers, we perform annealing over 2 hours in vacuum at temperatures $500 ^{\circ} \mathrm{C}$, $1100 ^{\circ} \mathrm{C}$ and $1300 ^{\circ} \mathrm{C}$ (Fig.~\ref{fig2}e). To prevent Si evaporation at $1300 ^{\circ} \mathrm{C}$, a thick layer ($2 \, \mathrm{\mu m}$) of UV resist is spin coated to the substrate and then baked at $180^{\circ} \mathrm{C}$ to remove the solvents (Fig.~\ref{fig2}f). The sample is placed in vacuum annealing chamber and pumped down to $2 \times 10^{-11} \, \mathrm{bar}$. It is then annealed at $1300^{\circ} \mathrm{C}$  for 2 hours and during the annealing process, the resist carbonizes forming the required carbon cap over the SiC substrate which prevents the migration of silicon to the surface of the substrate. After annealing, the carbon cap is removed with oxygen plasma etching at low power revealing the substrate surface, which is then cleaned with IPA to remove the loss material from the surface. 

\subsection{Spectral identification of the ClV defects}

\begin{figure}[t]
\centering\includegraphics[width=.47\textwidth]{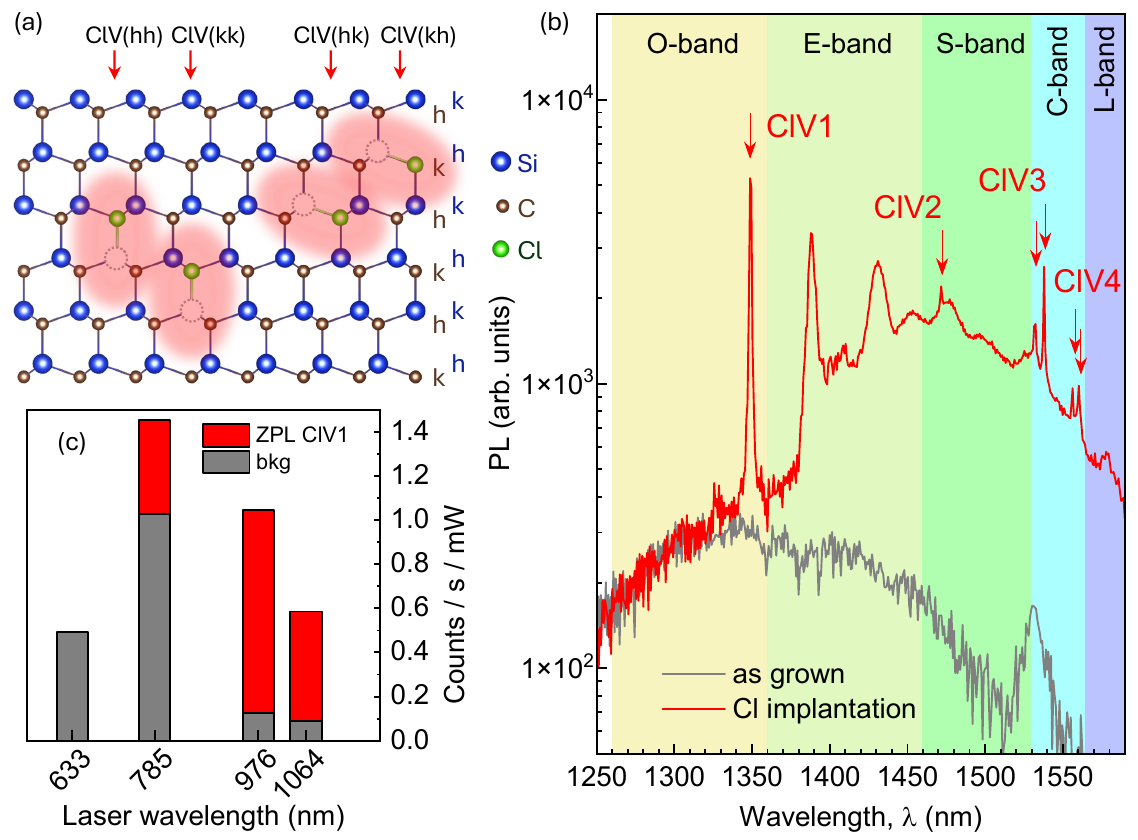}
\caption{Spectral fingerprins of ClV defects in 4H-SiC. (a) Schematic representation of the 4H-SiC lattice with different configuration of chlorine-vacancies in 4H-SiC ClV(hh), ClV(kk), ClV(hk) and ClV(kh).  (b) PL spectrum in Cl implanted 4H-SiC wafer with $E = 40 \, \mathrm{keV}$ under $976 \, \mathrm{nm}$ excitation at a temperauture $T = 7 \, \mathrm{K}$. The ZPLs from different configurations of ClV are labeled with arrows. (c) Normalized PL intensity of the ClV1 ZPL at $\lambda =1350 \, \mathrm{nm}$ and background (bkg) at $\lambda =1340 \, \mathrm{nm}$ for different excitation laser wavelength. }
\label{fig1}
\end{figure}

There are two non-equivalent lattice cites in 4H-SiC, which are historically labeled as h an k for both Si and C atoms in the lattice (Fig.~\ref{fig1}a). Correspondingly, there are four possible configurations for the ClV, where Cl occupies the C site due to its close atomic radius and bound to the Si vacancy (Table~\ref{ZPL_ClV}). 

\begin{table*}[t]
\caption{Configurations of the ClV defects in 4H-SiC: comparison DFT calculation \cite{10.1103/physrevb.108.224106} and our experiment. }
\label{ZPL_ClV}
\begin{center}
\begin{tabular}{c|c|c|c|c}
Crystal configuration & hh, on-axis & kk, on-axis & hk, off-axis & kh, off-axis \\
\hline
Symmetry & $C_{3v}$ & $C_{3v}$ &  $C_{1h}$ & $C_{1h}$ \\
ZPL, DFT theory (nm)  &  $1330$  & $1440$  & $1490$  & $1590$ \\
Notation   &  ClV1  & ClV2  & ClV3 & ClV4 \\
ZPL, experiment (nm)  &  $1350$  & $1472$  & $1532$, $1538$  & $1556$, $1561$ \\ 
Telecom window &  O-band  & S-band  & C-band  & C-band 
\end{tabular}
\end{center}
\label{default}
\end{table*}%

To identify different configuration of ClV defects, we compare the PL spectrum in the sample \#3 as grown and after Cl implantation with an energy $E = 40 \, \mathrm{keV}$ to a fluence $\Phi = 1 \times 10^{12} \, \mathrm{cm ^{-2}}$ followed by annealing at a temperature of  $1100 ^{\circ} \mathrm{C}$ over two hours (Fig.~\ref{fig1}b). There are spectrally-narrow lines, which appears only after Cl implantation. We associate them to the ZPLs  with chlorine-vacancy defects 
and label as ClV1, ClV2, ClV3 and ClV4 in Fig.~\ref{fig1}b. The spectrally-broad lines between the ClV1 and ClV2 ZPLs are ascribed to the PSBs. The theory predicts that the on-axis configurations ClV(hh) and ClV(kk) emit at shorter wavelengths $1330 \, \mathrm{nm}$ and $1440 \, \mathrm{nm}$  \cite{10.1103/physrevb.108.224106}. Therefore, we associate the ZPL CLV1 at $1350 \, \mathrm{nm}$ and  ZPL CLV2 at $1472 \, \mathrm{nm}$ to the hh and kk configurations, respectively, as schematically depicted in Fig.~\ref{fig1}a. The off-axis configurations hk and kh  of the ClV defects (Fig.~\ref{fig1}a) have lower symmetry ($C_{1h}$) compared to the on-axis configuration ($C_{3v}$), which may lead to the splitting of optical transition. Indeed, we observe that the ClV3 and ClV4 ZPLs are located in the telecom C-band ($1530 - 1565 \, \mathrm{nm}$), with the corresponding wavelengths listed in Table~\ref{ZPL_ClV}, and are split by about $6 \, \mathrm{nm}$ ($3.2 \, \mathrm{meV}$) and $5 \, \mathrm{nm}$ ($2.6 \, \mathrm{meV}$), respectively (Fig.~\ref{fig1}b). 

The DFT calculations predict that the optically active ClV in the telecom bands are positively charged and have spin ground state $S = 1$ \cite{10.1103/physrevb.108.224106}. The charge state can be verified by integrating the ClV defects into  p-i-n diodes  \cite{10.1103/pv13-vgcw} and the spin state can be determined experimentally using optically detected magnetic resonance (ODMR) \cite{10.1038/nphys2826}.  Both experiments  are beyond the scope of this work and planned in the future. 

\section{Optical properties of chlorine-based imitters in SiC}

\subsection{Spectral fingerprints}

We now concentrate on the optical properties of the ClV1 defect with the strongest ZPL in the telecom O-band ($1260 - 1360 \, \mathrm{nm}$). Figure~\ref{fig1}c shows the count rate of the ClV1 ZPL and the background for different excitation wavelengths, normalized to the excitation power. The most effective excitation is $ 976 \, \mathrm{nm}$, which is used throughout the remainder of this work. 

\begin{figure}[t]
\centering\includegraphics[width=.48\textwidth]{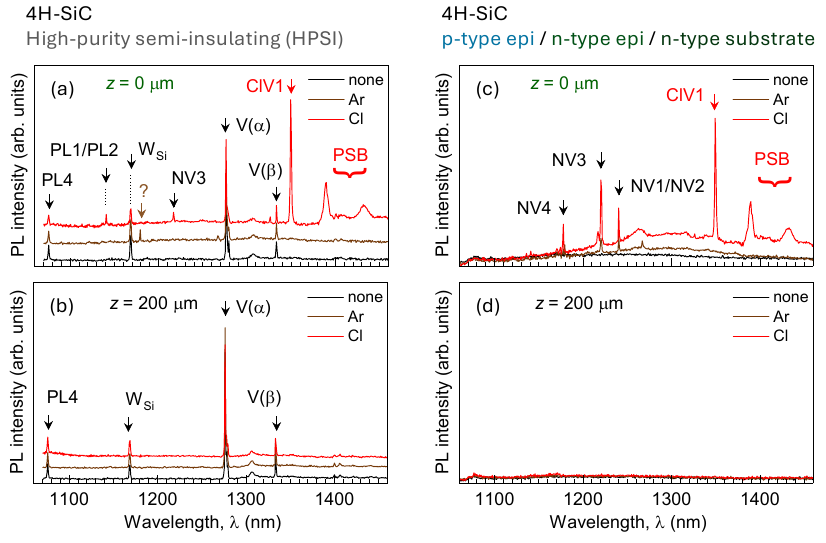}
\caption{Optical fingerprints of color centers in 4H-SiC. (a) PL spectra in HPSI 4H-SiC at a depth $z = 0 \, \mathrm{\mu m}$ after annealing at $1100 ^{\circ} \mathrm{C}$ over 2 hours without implantation compared to Ar and Cl implantation at an energy $E = 150 \, \mathrm{keV}$ and to a fluence $\Phi = 1 \times 10^{12} \, \mathrm{cm ^{-2}}$.  The vertical arrows indicate the ZPLs of divacanccies (PL4 and PL1/PL2), tungsten substituting silicon ($\mathrm{w_{Si}}$), nitrogen-vacancy NV3, vanadium in different lattice cites $\mathrm{V (\alpha)}$ and $\mathrm{V (\beta)}$ as well as of chlorine-vacancy  (ClV) and its possible PSB. (b) The same as (a) but measured at a depth $z = 200 \, \mathrm{\mu m}$ below the implanted surface.  (c) PL spectra in epetaxial 4H-SiC at a depth $z = 0 \, \mathrm{\mu m}$ after annealing at $1100 ^{\circ} \mathrm{C}$ over 2 hours without implantation compared to Ar and Cl implantation at an energy $E = 150 \, \mathrm{keV}$ and to a fluence $\Phi = 1 \times 10^{12} \, \mathrm{cm ^{-2}}$.  The vertical arrows indicate the ZPLs of nitrogen-vacancies NV4, NV3, NV1/NV2  as well as of chlorine-vacancy  (ClV) and its possible PSB. (b) The same as (a) but measured at a depth $z = 200 \, \mathrm{\mu m}$ below the implanted surface.    }
\label{fig3}
\end{figure}

To unambiguously attribute the newly observed color centers in 4H-SiC to chlorine incorporation, we conducted a series of control experiments using different wafers that underwent identical thermal treatments but were subjected to different ion implantations, namely chlorine, argon (the next element in the periodic table) and no implantation. Figures~\ref{fig3}a and b show PL spectra measured at the surface ($z = 0\,\mathrm{\mu m}$) and at a depth of $z = 200\,\mathrm{\mu m}$ below the surface, respectively, in the HPSI 4H-SiC wafer (sample \#1). At $z = 200\,\mathrm{\mu m}$, no discernible differences are observed between the Cl-, Ar-implanted and as-grown samples. This result is expected, as both the implanted ion profiles (Fig.~\ref{fig2}a) and the corresponding vacancy distributions (Fig.~\ref{fig2}b) are confined to a shallow region ($< 1\,\mathrm{\mu m}$) near the surface, which is much shallower than the $200\,\mathrm{\mu m}$ probing depth. Based on the spectral fingerprints of ZPLs, we identify in our HPSI 4H-SiC wafers the presence of various known color centers, including divacancies (PL4) \cite{10.1103/physrevb.111.165201}, tungsten-related centers ($\mathrm{W_{Si}}$) \cite{10.1557/mrc.2017.56} as well as  vanadium-related centers at the $k$-  and $h$-sites, labeled as  $\mathrm{V(\alpha)}$ and $\mathrm{V(\beta)}$, respectively \cite{10.1103/physrevapplied.12.014015}. 

In contrast, at the surface, the ClV1 ZPL and its expected phonon sideband are observed exclusively in the Cl-implanted sample. This experiment provides direct evidence that the newly observed lines are not caused by intrinsic defects, but instead originate specifically from chlorine incorporation.
 
We also repeat the same procedure in a 4H-SiC epi layer (sample \#2). No detectable PL is observed at a depth $z = 200\,\mathrm{\mu m}$, corresponding to the heavily doped $n$-type substrate (Fig.~\ref{fig3}c). At the surface ($z = 0\,\mathrm{\mu m}$), the ZPLs associated with the NV defects appear after Cl and Ar implantation (Fig.~\ref{fig3}d). They are likely formed at the boundary between the $p$-type epi layer and the $n$-type, nitrogen-doped buffer epi layer, due to  ion bombardment followed by nitrogen diffusion. All four possible configurations NV1--NV4 are clearly identified by their spectral fingerprints~\cite{10.1103/physrevb.94.060102, 10.1103/physrevlett.124.223601}. However, the ClV1 ZPL and its PSB are only observed after Cl implantation, supporting the conclusion drawn from sample~\#1.

\subsection{Optimization of the creation protocol}

\begin{figure}[t]
\centering\includegraphics[width=.44\textwidth]{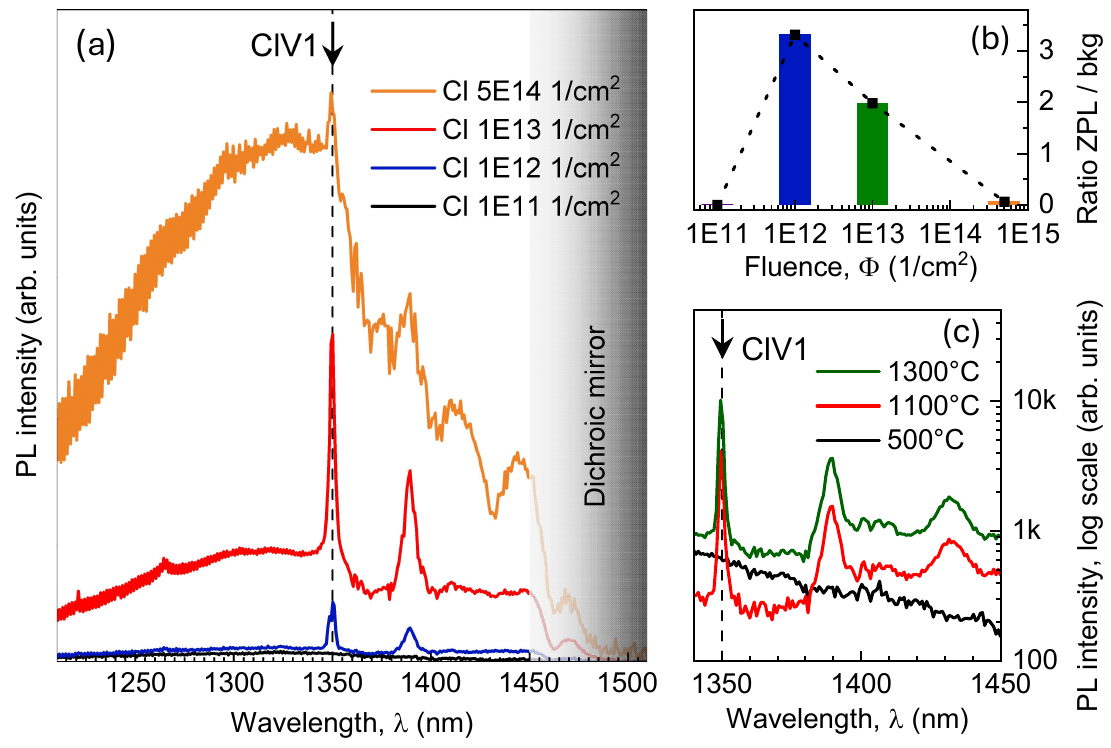}
\caption{Dependence on implantation fluence and annealing temperature. (a) PL spectrum at different implantation fluences for an annealing temperature of $1100 ^{\circ} \mathrm{C}$. The arrow indicates the spectral position of the ClV ZPL. The decrease in PL intensity within the shaded spectral region is attributed to reduced reflection of  the dichroic mirror. (b) Ratio ClV ZPL to the PL background for different implantation fluences $\Phi$. (c) PL spectra plotted on a logarithmic scale, measured after Cl implantation to a fluence $\Phi = 1 \times 10^{12} \, \mathrm{cm ^{-2}}$  and subsequent annealing at different temperatures. }
\label{fig4}
\end{figure}

Having proven the origin of the ClV color centers, we now turn to the optimization of their creation protocol. Figure~\ref{fig4}a shows PL spectra in sample~\#1 for different chlorine implantation fluences, followed by annealing at $1100\,^{\circ}\mathrm{C}$. No ClV1 signal is detected for the implantation fluence $\Phi = 1 \times 10^{11}\,\mathrm{cm^{-2}}$. This and lower fluences are promising for the isolation of single ClV defects, which is beyond the scope of this work. At $\Phi = 1 \times 10^{12}\,\mathrm{cm^{-2}}$, the ClV1 ZPL is clearly visible with a low background. For higher fluences, we observe an increase in the ClV1 ZPL intensity, accompanied by a rapid rise in background due to PSBs from other defects. At the highest fluence $\Phi = 5 \times 10^{14}\,\mathrm{cm^{-2}}$ this background dominates the PL spectrum. The ratio of the ClV1 ZPL at a wavelength of $1350\,\mathrm{nm}$ to the background measured at $1340\,\mathrm{nm}$ is summarized in Fig.~\ref{fig4}b for different fluences. According to this fluence dependence, $\Phi = 1 \times 10^{12}\,\mathrm{cm^{-2}}$ yields the highest ClV1 ZPL-to-background ratio, identifying it as the optimal condition for ClV ensemble creation under the given annealing parameters.

Figure~\ref{fig4}c presents the effect of the annealing temperature on the formation of the ClV defects after implantation. We find that annealing  by $500\,^{\circ}\mathrm{C}$ over 2 hours does not lead to the formation of ClV defects. The ClV1 ZPL is observed at the annealing temperature $1100\,^{\circ}\mathrm{C}$ and its intensity doubled if the annealing temperature increases to $1300\,^{\circ}\mathrm{C}$. 

\begin{figure}[t]
\centering\includegraphics[width=.47\textwidth]{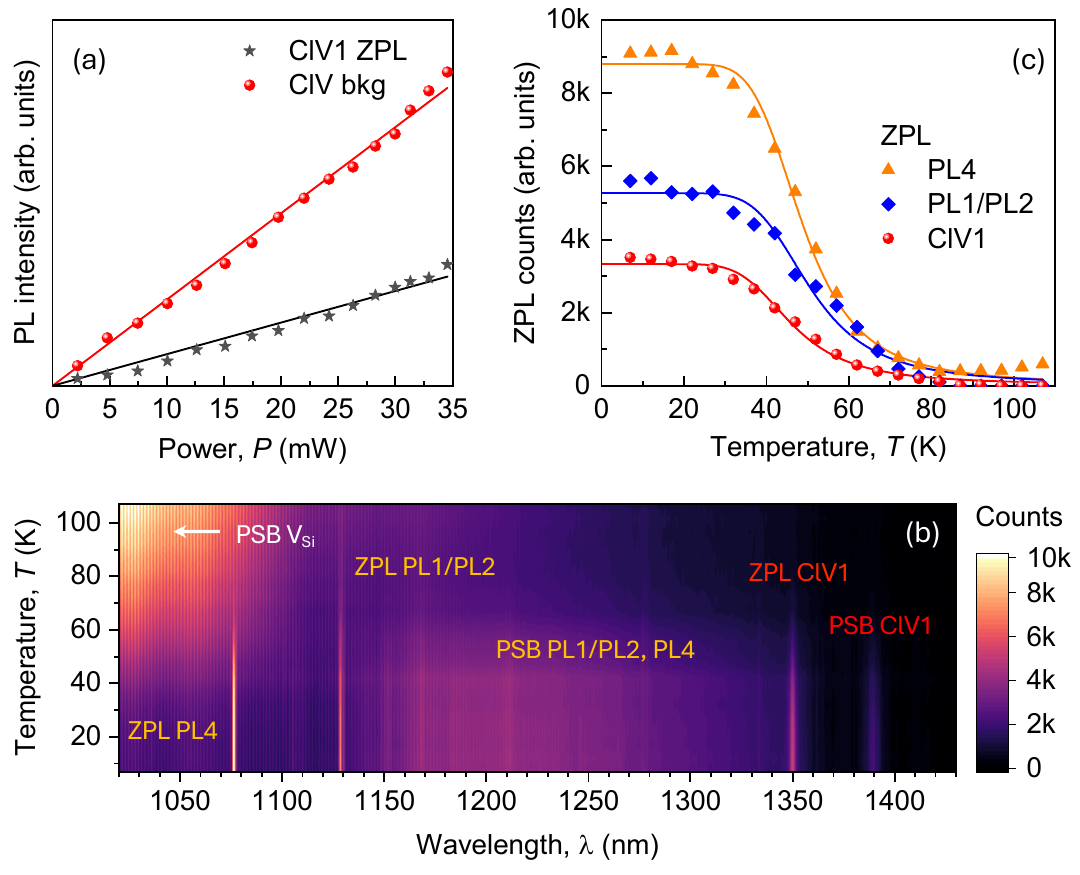}
\caption{PL power and temperature dependence. (a) Intensity of the ClV ZPL and backgroud as a function of the laser power $P$ with the excitation  wavelength of $976 \, \mathrm{nm}$. The solid lines are linear fits. (b) 2D color plot showing temperature evolution of the PL spectrum. (c) Temperature dependence of the ZPL from divacancies PL4 and PL1/PL2 together with the ClV. The solid lines are fit to Eq.~(\ref{Activation_Energy}).   }
\label{fig5}
\end{figure}

The potential for further optimization of the annealing protocol is indirectly supported by the power dependence shown in Fig.~\ref{fig5}a. We observe a linear increase in the ClV1 ZPL intensity with excitation power up to $35\,\mathrm{mW}$ without any indication of saturation. This suggests the presence of non-radiative recombination channels, likely due to incomplete lattice recovery during annealing after ion implantation damage.

Finally, we analyze the temperature dependence of the PL spectrum shown in Fig.~\ref{fig5}b. The PL from divacancies (PL1/PL2, PL4) and ClV defects decreases by about one order of magnitude as the temperature increases from $7\,\mathrm{K}$ to $80\,\mathrm{K}$, accompanied by the emergence of the PSB tail from $\mathrm{V_{Si}}$ in the short-wavelength region of the spectrum. The temperature dependence of the ZPL intensity $I_{\mathrm{ZPL}}$ from ClV1, in comparison with that from divacancies PL4 and PL1/PL2, is summarized in Fig.~\ref{fig5}c. The data are fitted using the Arrhenius law 
\begin{equation}\label{Activation_Energy}
I_{\mathrm{ZPL}} = \frac{I_0}{1 + B \, e^{-\frac{E_a}{k_B T}}} \,.
\end{equation}
The temperature dependence of the ClV1 ZPL closely follows that of the divacancies, with an extracted activation energy of $E_a = 25 \pm 11\,\mathrm{meV}$. A detailed understanding of the energetic structure of the ClV defect is required for a proper interpretation of this activation energy.

\section{Conclusion}

We have experimentally observed chlorine-vacancy (ClV) color centers in 4H-SiC, establishing a new class of optically active defects emitting across the entire fiber-optic telecom range. The ClV centers exhibit ZPLs in the strategically important O- and C-bands, the latter corresponding to the lowest transmission loss in optical fibers. We have optimized the creation protocol for ClV ensembles, identifying an implantation fluence of $\Phi = 1 \times 10^{12}\,\mathrm{cm^{-2}}$ and annealing at $1300\,^{\circ}\mathrm{C}$ as optimal conditions. Annealing at even higher temperatures, which are not accessible in the current experiments, may further enhance the signal-to-background ratio. The ClV ZPL shows negligible reduction in intensity with increasing temperature up to $30\,\mathrm{K}$. From an Arrhenius fit of the temperature dependence, we estimate an activation energy $E_a = 25 \pm 11\,\mathrm{meV}$. These results demonstrate the thermal stability and robustness of the ClV optical transition.

DFT-based analysis predicts a non-zero spin in the ground state of the ClV defect \cite{10.1103/physrevb.108.224106}, making it a promising candidate for spin-photon interfaces and quantum memories operating at telecom wavelengths. A natural next step is the experimental investigation of its spin properties via optically detected magnetic resonance (ODMR). Furthermore, isolating single ClV emitters and integrating them into photonic cavities can potentially enable bright single-photon sources and form the basis for quantum repeaters.

In summary, the experimental identification of ClV centers may pave the way for future advances in telecom-band quantum technologies based on the technologically mature and CMOS-compatible SiC platform.

\section*{Acknowledgments}
This work was funded by the European Union under project 101186889 QuSPARC. 
Support from the Ion Beam Center (IBC) at Helmholtz-Zentrum Dresden-Rossendorf (HZDR) is also acknowledged.


\bibliography{SiC_ClV_references} 

\end{document}